\documentclass[preprint]{aastex}

\shorttitle{THE EFFECT OF HELIUM-ENHANCED STELLAR POPULATION ON THE ULTRAVIOLET-UPTURN PHENOMENON OF EARLY-TYPE GALAXIES}
\shortauthors{Chung et al.}

\begin{document}

\title{THE EFFECT OF HELIUM-ENHANCED STELLAR POPULATIONS ON THE ULTRAVIOLET-UPTURN PHENOMENON OF EARLY-TYPE GALAXIES}

\author{CHUL CHUNG, SUK-JIN YOON, AND YOUNG-WOOK LEE} 
\affil{Center for Galaxy Evolution Research and Department of Astronomy, Yonsei University, Seoul 120-749, Korea; ywlee2@yonsei.ac.kr}

\begin{abstract}

Recent observations and modeling of globular clusters with multiple populations strongly indicate the presence of super helium-rich subpopulations in old stellar systems.
Motivated by this, we have constructed new population synthesis models with and without helium-enhanced subpopulations to investigate their impact on the UV-upturn phenomenon of quiescent early-type galaxies.
We find that our models with helium-enhanced subpopulations can naturally reproduce the strong UV-upturns observed in giant elliptical galaxies assuming an age similar to that of old globular clusters in the Milky Way.
The major source of far-UV (FUV) flux, in this model, is relatively metal-poor and helium-enhanced hot horizontal branch stars and their progeny.
The \citet{1988ApJ...328..440B} relation of the $FUV-V$ color with metallicity is also explained either by the variation of the fraction of helium-enhanced subpopulations or by the spread in mean age of stellar populations in early-type galaxies.

\end{abstract}


\section{INTRODUCTION}

It is well established that the far ultraviolet (FUV) flux (``UV-upturn'') observed in the nearby quiescent early-type galaxies (ETGs) originates from a minority population of hot horizontal-branch (HB) stars and their progeny \citep[see, e.g.,][]{1990ApJ...364...35G, 1999ARA&A..37..603O, 2000ApJ...532..308B}.
Recently, a significant progress has been made as to the origin of these hot HB stars in old stellar systems. This new theory is based on the recent observations and modeling of globular clusters (GCs) with extended HB, such as $\omega$~Cen and NGC 2808, where the multiple main-sequences and hot HBs in these clusters can only be explained by the presence of super helium-rich subpopulations \citep{2004ApJ...611..871D, 2004ApJ...612L..25N, Lee05, 2005ApJ...621..777P, 2007ApJ...661L..53P}.

The origin of this helium enhancement is most likely due to the pollution from the intermediate-mass asymptotic giant branch stars and/or fast-rotating massive stars, or due to the enrichment by supernovae \citep{2008MNRAS.385.2034V, 2007A&A...464.1029D, 2005ApJ...621..777P, 2009Natur.462..480L}. 
Therefore, the likely presence of super-helium-rich subpopulations and the resulting hot HB stars in old stellar systems deserve further investigation, as they could be a major source of the FUV flux in quiescent elliptical galaxies. 
The purpose of this Letter is to report our first result on the UV upturn phenomenon predicted from the Yonsei evolutionary population synthesis models with helium-enhanced subpopulations. 

\section{POPULATION SYNTHESIS MODELS}

The models presented in this paper were constructed using the Yonsei Evolutionary Population Synthesis (YEPS) code.
The readers are referred to \citet{Park97}, \citet{Lee00}, and Chung et al. (2011, {\it in prep.}) for the details of model construction.
In order to investigate the effect of helium-enhanced stellar population on the UV-upturn, we have used the most up-to-date Yonsei-Yale ($Y^2$) stellar isochrones and HB evolutionary tracks (Han et al. 2011, {\it in prep.}) with different values of helium abundance ($Y = 0.23$, 0.33, and 0.38).
We apply the same \citet{1977A&A....57..395R} mass-loss parameter ($\eta$) and the enhancement in $\alpha$-elements ($[\alpha/Fe]$) for both helium-enhanced and normal helium populations (see Table 1).

Figure~1 shows an example of our model HR diagrams and corresponding spectral energy distributions (SEDs) of simple stellar population (SSP) for normal helium and helium-enhanced populations.
Since helium-rich stars evolve faster than helium-poor stars, helium-rich stars have lower masses at given age. 
This effect has a striking difference on the HB stage (see left panels). 
The mean temperature of HB stars in helium-enhanced ($Y=0.33$) population is $\sim$~11,500~K greater than that of the normal helium ($Y=0.23$) population at a given metallicity ($[Fe/H]=-0.9$) and age (11~Gyr).
Consequently, the helium-enhanced models show extremely strong far-UV flux compared to the normal helium models.

In order to include this effect of the helium-enhanced population into the composite SED for a model elliptical galaxy, we have employed following procedures in the model construction.
First, helium abundances of helium-enhanced subpopulations were chosen to be $Y = 0.33$ and 0.38, as these values are roughly required to reproduce the extreme blue HB stars observed in the Milky Way GCs (\citealt{2004ApJ...611..871D, 2008MNRAS.390..693D}; \citealt{Lee05}; \citealt{2005ApJ...621..777P}; Joo \& Lee 2011, {\it in prep.}).
In our modeling, populations having these two helium abundances were mixed half and half.
Second, we have used mean $FUV-V$ colors of UV-strong GCs in M87 \citep{2006AJ....131..866S} to set the fraction of helium-enhanced subpopulations at given metallicity (see Figure~2a).
Figure~2b shows the fraction of helium-enhanced subpopulations as a function of metallicity, which was adopted in our model with helium-enhanced subpopulations to reproduce the observed trend of $FUV-V$ color with metallicity in M87 GCs.
Figure~2c displays the contribution of helium-enhanced subpopulations in the metallicity distribution function (MDF) used in our construction of the composite stellar population model for giant elliptical galaxies.
This MDF was adopted from the simple chemical evolution model of \citet{1997A&A...320...41K}.

According to the list of \citet{2007ApJ...661L..49L}, about 30\% of the Milky Way GCs have extended HB, each of which contains about 30\% of helium-enhanced subpopulation \citep{Lee05, 2005ApJ...621..777P, 2007ApJ...661L..53P, Yoon08, Han09, 2010AJ....140..631B}, so about 9\% of stellar population in the Milky Way GCs are assumed to be helium-enhanced population.
In our composite model for giant elliptical galaxies, the number fraction of helium-enhanced subpopulations is about $\sim$ 11\%, which agrees roughly with that estimated in the Milky Way GCs.
This fraction of helium-enhanced subpopulations is reduced to $\sim$~6.4\% when only hot enough ($T_{\rm eff} \geq 20,000$~K) blue HB stars are counted, which are actually responsible for the far-UV flux in our model giant elliptical galaxies.\footnote{For the dwarf elliptical galaxy like M32, where the observed UV upturn is weak ($FUV-V$~=~7.3 in AB magnitude system), all of the HB population is predicted to be cooler than 20,000~K. 
\citet{2000ApJ...532..308B} found from their HST/STIS photometry of M32 that stars passing through the HB at $T_{\rm eff} \geq 8,500$~K comprise only a small fraction (about 7\%) of the total HB population. This fraction is roughly consistent with that ($\sim$~5.5\%) predicted from our models for M32.}

\section{COMPARISON WITH OBSERVATIONS}

Figure~3 presents our composite models constructed with and without helium-enhanced subpopulations compared with the SEDs observed by {\it IUE} satellite for two giant elliptical galaxies NGC~4552 and NGC~4649.
In our models, the value of mean metallicity ($\left<[Fe/H]\right>$) was chosen so that they can reproduce the observed Mg~{\it b} ($\approx$~4.6~{\AA}) index of theses galaxies (see also Figure~4), assuming that $[\alpha/Fe]$ is 0.3 in giant elliptical galaxies \citep{1992ApJ...398...69W, 2005ApJ...621..673T, 2009ApJS..182..216K}.
The parameters adopted in our models presented in Figure~3 are listed in Table 1.
It is clear from Figure~3 that our models with helium-enhanced subpopulations can naturally reproduce not only the strong far-UV upturns but also the 2500~{\AA} dips observed in NGC~4552 and NGC~4649, while the models without helium-enhanced subpopulations fail to reproduce the far-UV upturns, unless the age of underlying stellar population is increased to $\sim$~16~Gyrs \citep[][]{Park97, 1999ApJ...513..128Y}.
Therefore, the minority population ($\sim$~6.4\%) of helium-enhanced hot HB stars and their progeny is responsible for the observed far-UV flux in this scenario.\footnote{Most of the helium-rich HB stars in our models are too hot to have any significant effect on metal lines, such as Ca II H + H$\epsilon$/Ca II K index. 
The Ca II H + H$\epsilon$/Ca II K index of our model for giant elliptical galaxy (presented in Figure 3) is 1.15, which is consistent with the observation ($\sim$~1.16; \citealt{1985AJ.....90.1927R}) to within the error ($\pm$~0.03).}

Our composite models with helium-enhanced subpopulations predict that about 85\% of the far-UV flux comes from the metal-poor side ({$[Fe/H] \leq 0.0$}) of the MDF.
In this regard, this model is qualitatively similar to the ``metal-poor HB model'' of \citet{Park97}, but it does not need to invoke unrealistically old ages ($\sim$~16~Gyr) as in the model of \citet[][see also \citealt{1999ApJ...513..128Y}]{Park97}.
Our composite models with helium-enhanced subpopulations can reproduce the UV-upturn at the mean age of 11~Gyrs, in the age scale where the inner halo GCs of the Milky Way is 12~Gyrs old.
If the age of the oldest GCs in the Milky Way is more like 13~Gyrs \citep{2010ApJ...708..698D}, our age for the giant elliptical galaxies would also increase by 1~Gyr. 
These ages for giant elliptical galaxies (11~-~12~Gyrs) are consistent with the value estimated from the Balmer line indices to within the error \citep{Trag00, 2005ApJ...621..673T, 2008MNRAS.386..715T, 2009ApJ...693..486G}.

In Figure~4, we have plotted the Mg~$b$ line strength against $FUV-V$ color for the sample of quiescent ETGs from \citet{2011MNRAS.414.1887B}.
Superposed grids are our composite models with (blue) and without (red) helium-enhanced subpopulations.\footnote{Our models were constructed with $[\alpha/Fe]=0.3$, while some galaxies in Figure~4 have lower values of $[\alpha/Fe]$.
If $[\alpha/Fe]=0.1$, the $FUV-V$ color gets redder by small amount ($\sim$~0.1), while the Mg~$b$ index is more affected (decreases by $\sim$~0.4~\AA). 
Therefore, the model grids in Figure 4 shift mostly downwards as $[\alpha/Fe]$ decreases, and this should have only little effect on the predicted age spread among sample galaxies.}
The observed ETGs show clear correlation between $FUV-V$ and Mg~$b$ that is analogous to the \citet{1988ApJ...328..440B} relation.
Our composite models predict that the strength of UV-upturn is controlled by the fraction of helium-enhanced subpopulation, the mean age, and the mean metallicity of underlying stellar population.
Among these three factors, the fraction of helium-enhanced subpopulation appears to be the most effective factor determining the strength of UV-upturn.
For example, our 11~Gyr models with and without helium-enhanced subpopulations would explain most of the color spread in $FUV-V$.
If all ETGs in this sample, however, contain similar fraction of helium-enhanced subpopulations, an age spread spanning $\sim$~6~Gyrs would be required to reproduce the observed spread in $FUV-V$ color, in the sense that bluer galaxies are older.
This is because the mean temperature of HB stars decreases as the age of stellar population gets younger \citep[see ][]{LDZ94, Park97}.
An age spread of this magnitude is also indicated from the Balmer line dating of ETGs \citep{Trag00, 2005ApJ...621..673T, 2008MNRAS.386..715T, 2009ApJ...693..486G}, which suggests that some age spread is indeed responsible for the spread in $FUV-V$ color.
The observed fading of FUV flux with look-back time for the bright cluster elliptical galaxies in $z < 0.3$ \citep{Ree07} also supports this possibility.
Note from Figure~4 that our models both with and without helium-enhanced subpopulations predict fading of FUV flux with increasing look-back time.
\section{DISCUSSION}

We have demonstrated that the presence of relatively metal-poor and helium-enhanced subpopulations in ETGs can naturally reproduce the observed UV-upturn phenomenon, without invoking unrealistically old ages. 
Although the presence of helium-enriched stars appears to solve many of the problems associated with the UV-upturn phenomenon, it is important to note that the origin of this helium enhancement is not fully understood yet. 
Consequently, it is not clear yet whether the proposed mechanisms to produce helium-enhanced stars in GCs will also be important in the galactic scales. 
Nevertheless, several lines of evidence do suggest that a minority population of helium-enhanced stars would be present also in ETGs. 
First, the two well defined remaining local building blocks in the Milky Way, the $\omega$~Cen and the Sagittarius dwarf galaxy (including M54), are all characterized by very extended HB with helium-enhanced extreme blue HB stars (\citealt{Lee05}; \citealt{2005ApJ...621..777P}; \citealt{2010AJ....140..631B}; \citealt{2007ApJ...667L..57S}; \citealt{2008MNRAS.390..693D}; Joo \& Lee 2011, {\it in prep.}). 
Second, the orbital kinematics of Milky Way GCs with extended HB, which is indicating the presence of helium-enhanced subpopulation, are distinct from normal GCs, and are fully consistent with the hypothesis that they are the remaining relics of the early building blocks predicted in the hierarchical merging paradigm of galaxy formation \citep{2007ApJ...661L..49L, 2007MNRAS.382L..87B}. 
Finally, as discussed above, when the observed trend of $FUV-V$ color with metallicity for the UV-strong GCs (presumably with the extended HB) in the giant elliptical galaxy M87 is combined with the expected MDF of a model giant elliptical galaxy, the $FUV-V$ color similar to that observed in giant elliptical galaxies is reproduced. 
This is again consistent with the building block origin for these GCs, according to which the helium-enhanced subpopulations in ETGs would have been supplied by these early galaxy building blocks.

If normal ETGs are prevailed by helium-enhanced populations, as suggested in this Letter, they would have impacts not only on UV, but also on other wavelength regime affected by relatively hot blue HB stars, such as Balmer absorption lines \citep{Lee00}.
In this respect, it is important to note that \citet{2006ApJ...651L..93S} found that their models (without helium-enhanced subpopulation) of passive evolution can not reproduce the unexpectedly strong Balmer line indices of ETGs at $z\approx0.1$ and $z\approx0.9$.
It would be interesting to see how our models with helium-enhanced subpopulations are matched with these observations.
The well known discrepancy between the H$\beta$ and higher order Balmer lines (``Balmer mismatch''; \citealt{Schi07}) is also an important issue, for which the helium-enhanced populations could play a role.
Our forthcoming paper (Chung et al. 2011, {\it in prep.}) will discuss these issues in more detail, together with their connection with the UV-upturn phenomenon.

\acknowledgments

We thank the referee for a number of helpful suggestions.
Support for this work was provided by the National Research Foundation of Korea to the Center for Galaxy Evolution Research.
SJY acknowledges support from Mid-career Researcher Program (No.~2009-0080851) and Basic Science Research Program (No.~2009-0086824) through the NRF of Korea, and support from the KASI Research Fund 2011.

\clearpage

\begin{figure}
\includegraphics[angle=-90,scale=1.2]{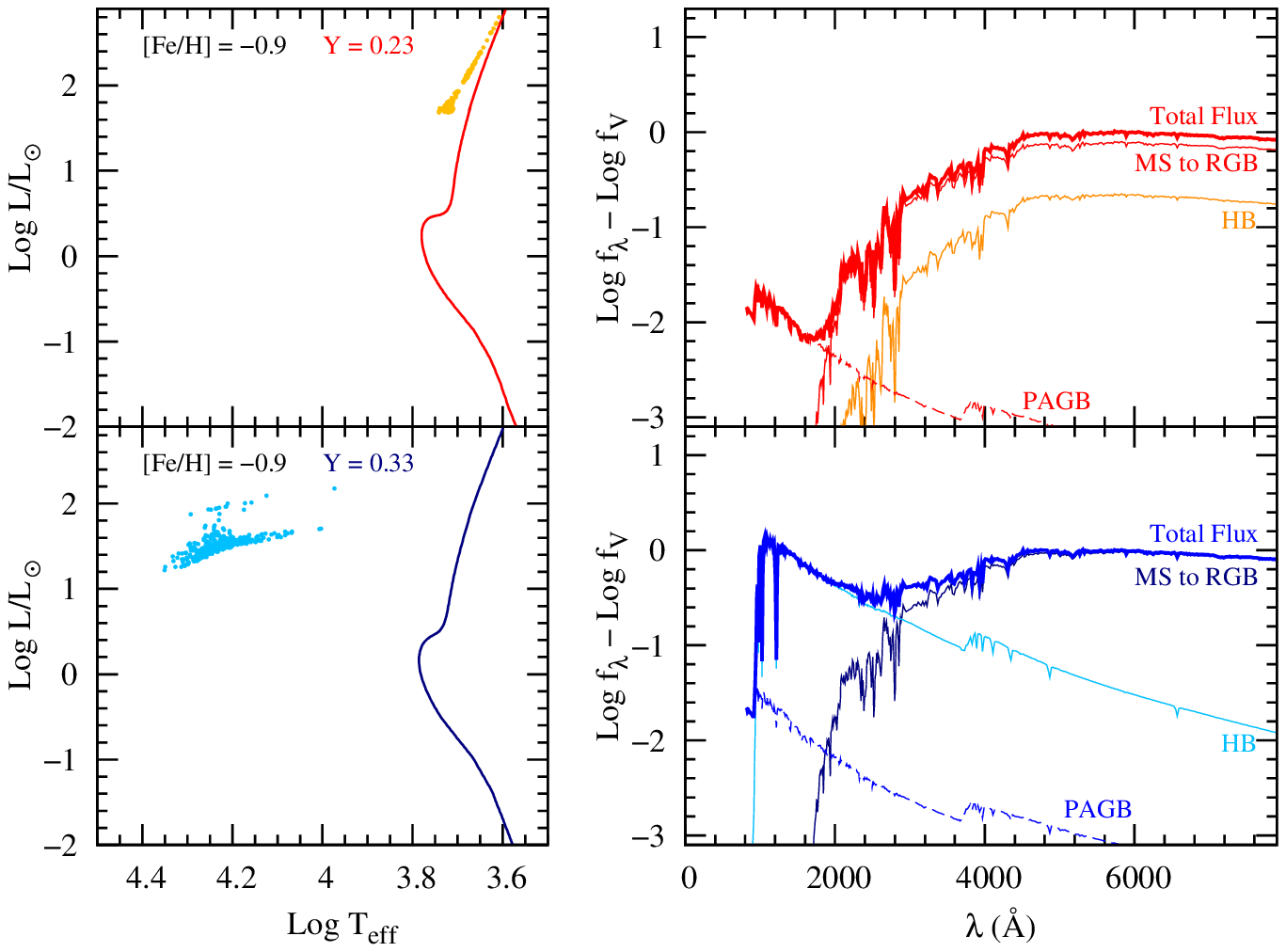}
\caption{
The synthetic HR diagrams and corresponding spectral energy distributions for simple stellar populations.
Upper panels are models for normal helium abundance ($Y = 0.23$), while bottom panels are for helium-enhanced ($Y = 0.33$) population.
The abbreviations MS, RGB, HB, and PAGB refer the flux contributions from main-sequence, red giant branch, horizontal branch (including post-HB progeny), and post asymptotic giant branch stars, respectively.
}
\end{figure}

\clearpage

\begin{figure}
\includegraphics[angle=-90,scale=0.61]{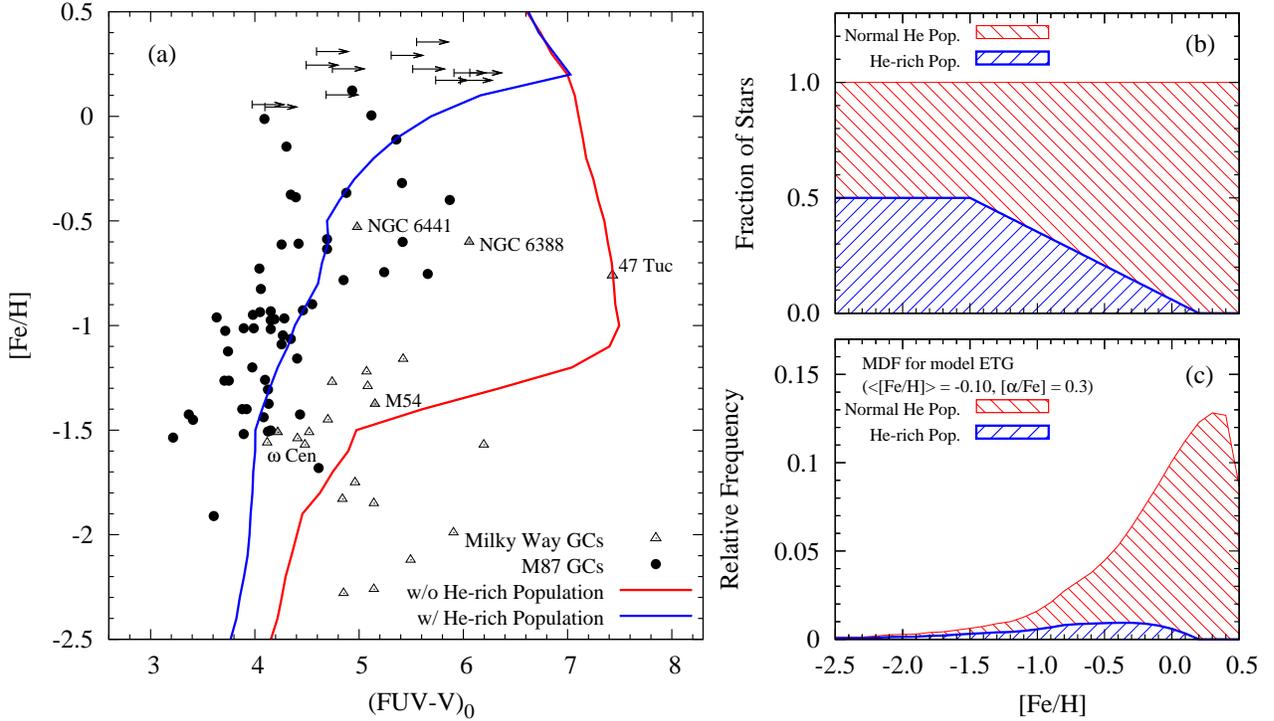}
\caption{\small ({\it a}) Comparison of our models ({\it solid lines}) with GCs in M87 ({\it circles}) and the Milky Way ({\it triangles}).
The data are from \citet{2006AJ....131..866S} for GCs in M87 and from \citet{1995ApJ...442..105D} for GCs in the Milky Way, and the $FUV-V$ color is in the AB magnitude system.
Bared arrows are upper limits for the metal-rich GCs in M87.
The red line is a model with only normal helium population ($Y=0.23$; $t=12$~Gyr), while the blue line is a model when the helium-enhanced subpopulations ($Y=0.33$~+~0.38; $t=11$~Gyr) are added to the normal helium population ($Y=0.23$; $t=11$ Gyr) using the fraction in Figure~2b.
Some well-known Milky Way GCs with extended HB ($\omega$~Cen, M54, NGC~6388, and NGC~6441), together with 47~Tuc, are labeled.
$[Fe/H]$ values for $\omega$~Cen and M54 are weighted mean values of multiple populations in these GCs.
({\it b}) The fraction of helium-enhanced population calibrated to reproduce $FUV-V$ colors of UV-strong GCs in M87 at given metallicity.
In our models (the blue line in Figure~2a), helium-enhanced subpopulations are added to the normal helium populations with this relation.
({\it c}) The metallicity distribution function \citep{1997A&A...320...41K} adopted in our composite model for giant elliptical galaxy and the contribution from helium-enhanced subpopulations.
Considering the metallicity spread of composite model peaked at $[Fe/H]=0.3$, the actual contribution from helium-enhanced subpopulation is only $\sim$~11\%. 
}

\end{figure}

\clearpage

\begin{figure}
\includegraphics[angle=-90,scale=1.1]{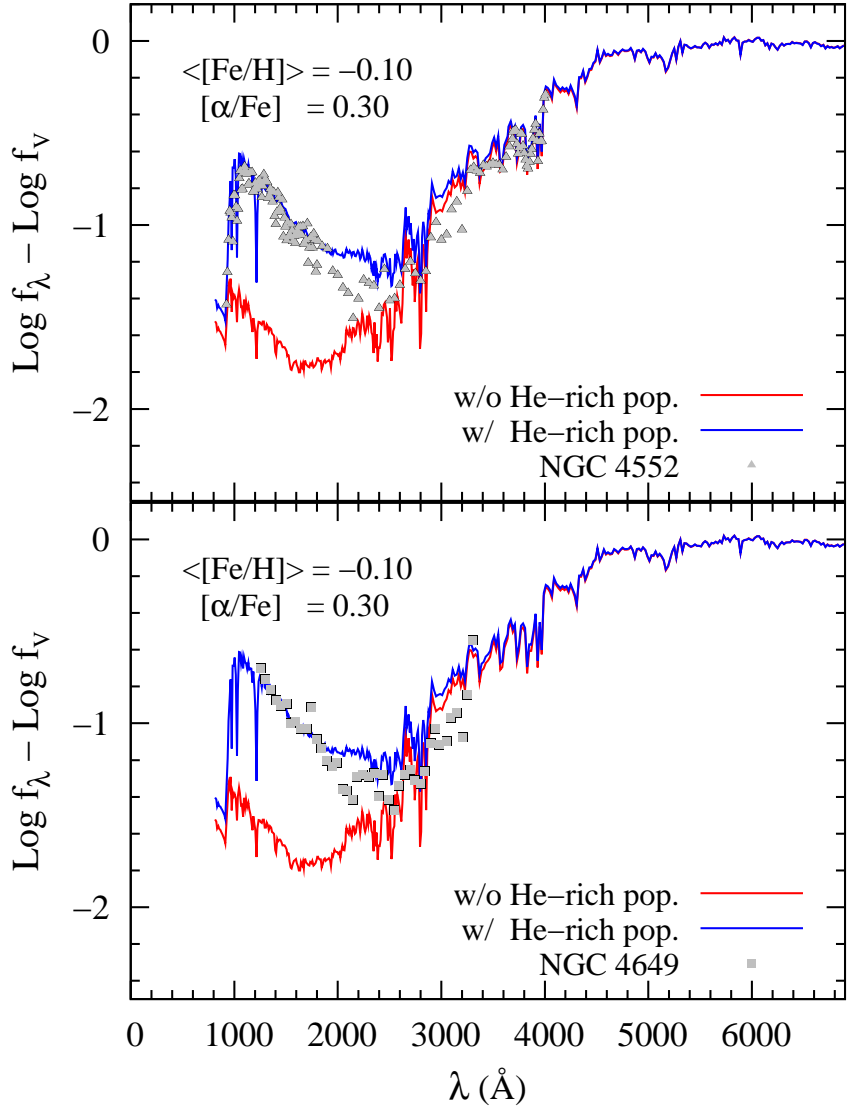}
\caption{The comparison of observed SEDs of NGC~4552 and NGC~4649 (data from \citealt{1988ApJ...328..440B} and \citealt{1998ApJ...492..480Y}) with our composite models. 
The composite models presented here are based on the series of SSP models such as those displayed in Figure~1, with the metallicity distribution function in Figure~2c.
Our models with helium-enhanced subpopulation can reproduce the observed far-UV upturns at the age of 11~Gyrs.
}
\end{figure}

\clearpage

\begin{figure}
\includegraphics[angle=0,scale=1.4]{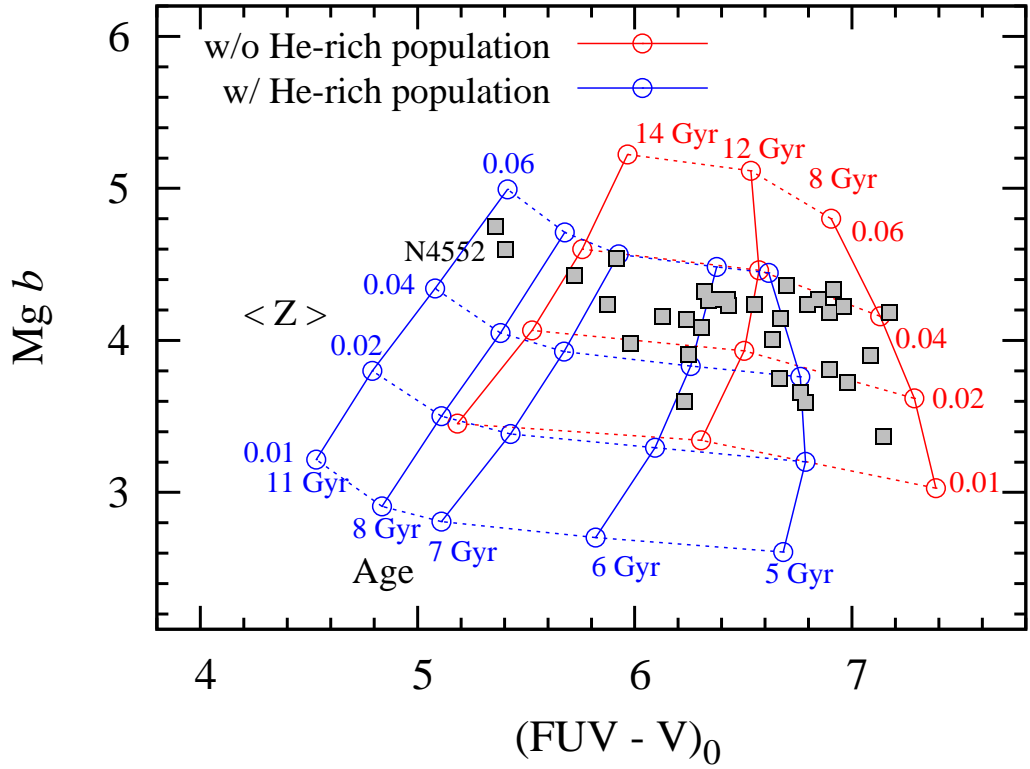}
\caption{The $(FUV-V)_0$ color vs. Mg~{\it b} correlation for the sample of quiescent early-type galaxies having H$\beta$ $\le$ 1.9 {\AA} from \citet{2011MNRAS.414.1887B}.
The blue and red lines are our composite models with and without helium-enhanced subpopulation, respectively, where
the iso-metal lines are connected by dotted lines.
The values of $\left< Z \right>$ and age (Gyr) are indicated.}
\end{figure}

\clearpage

\clearpage

\begin{deluxetable}{lcc}
\tabletypesize{\scriptsize}
\tablewidth{0pt}
\tablecaption{\label{tab:table1} INPUT PARAMETERS ADOPTED IN OUR COMPOSITE STELLAR POPULATION MODEL FOR NGC~4552}
\tablehead{
\colhead{Parameters} &\colhead{Helium-enhanced population} &\colhead{Normal Helium population}}
\startdata
Initial mass function&Salpeter ($x=1.35$)&Salpeter ($x=1.35$) \\
$\alpha$-elements enhancement, [$\alpha$/Fe]& 0.3& 0.3 \\
HB mass dispersion, ${{\sigma_{}}_{M}}$ ($M_{\odot}$)&0.015&0.015\\
Reimers' mass-loss parameter, $\eta$&0.63&0.63\\
Absolute age, $t$ (Gyr)&11.0&11.0\\
Mean metallicity, $\left<[Fe/H]\right>$&-0.76&-0.02\\
Helium abundance, $Y$ &0.33~+~0.38&0.23\\
Population ratio (\%)&11&89\\

\enddata
\end{deluxetable}


\begin{thebibliography}{}

\bibitem[Bekki et al.(2007)]{2007MNRAS.382L..87B} Bekki, K., Yahagi, H., 
Nagashima, M., \& Forbes, D.~A.\ 2007, \mnras, 382, L87
\bibitem[Bellini et al.(2010)]{2010AJ....140..631B} Bellini, A., Bedin, 
L.~R., Piotto, G., Milone, A.~P., Marino, A.~F., 
\& Villanova, S.\ 2010, \aj, 140, 631
\bibitem[Brown et al.(2000)]{2000ApJ...532..308B} Brown, T.~M., Bowers, 
C.~W., Kimble, R.~A., Sweigart, A.~V., 
\& Ferguson, H.~C.\ 2000, \apj, 532, 308
\bibitem[Bureau et al.(2011)]{2011MNRAS.414.1887B} Bureau, M., et al.\ 
2011, \mnras, 414, 1887
\bibitem[Burstein et al.(1988)]{1988ApJ...328..440B} Burstein, D., Bertola, 
F., Buson, L.~M., Faber, S.~M., \& Lauer, T.~R.\ 1988, \apj, 328, 440
\bibitem[D'Antona 
\& Caloi(2004)]{2004ApJ...611..871D} D'Antona, F., \& Caloi, V.\ 2004, \apj, 611, 871
\bibitem[D'Antona 
\& Caloi(2008)]{2008MNRAS.390..693D} D'Antona, F., \& Caloi, V.\ 2008, \mnras, 390, 693
\bibitem[Decressin et 
al.(2007)]{2007A&A...464.1029D} Decressin, T., Meynet, G., Charbonnel, C., Prantzos, N., \& Ekstr{\"o}m, S.\ 2007, \aap, 464, 1029
\bibitem[Dorman et al.(1995)]{1995ApJ...442..105D} Dorman, B., O'Connell, 
R.~W., \& Rood, R.~T.\ 1995, \apj, 442, 105
\bibitem[Dotter et al.(2010)]{2010ApJ...708..698D} Dotter, A., et al.\ 
2010, \apj, 708, 698
\bibitem[Graves et al.(2009)]{2009ApJ...693..486G} Graves, G.~J., Faber, 
S.~M., \& Schiavon, R.~P.\ 2009, \apj, 693, 486
\bibitem[Greggio 
\& Renzini(1990)]{1990ApJ...364...35G} Greggio, L., \& Renzini, A.\ 1990, \apj, 364, 35
\bibitem[Han et al.(2009)]{Han09} Han, S.-I., Lee, Y.-W., Joo, S.-J., Sohn, S.~T., Yoon, S.-J., Kim, H.-S., \& Lee, J.-W.\ 2009, \apjl, 707, L190 
\bibitem[Kodama 
\& Arimoto(1997)]{1997A&A...320...41K} Kodama, T., \& Arimoto, N.\ 1997, \aap, 320, 41
\bibitem[Kormendy et al.(2009)]{2009ApJS..182..216K} Kormendy, J., Fisher, 
D.~B., Cornell, M.~E., \& Bender, R.\ 2009, \apjs, 182, 216
\bibitem[Lee et al.(2000)]              {Lee00}Lee, H.-c., Yoon, S.-J., \& Lee, Y.-W.\ 2000, \aj, 120, 998
\bibitem[Lee et al.(2009)]{2009Natur.462..480L} Lee, J.-W., Kang, Y.-W., Lee, J., \& Lee, Y.-W.\ 2009, \nat, 462, 480
\bibitem[Lee et al.(1994)]{LDZ94}Lee Y.-W., Demarque P., Zinn, R., 1994, \apj, 423, 248
\bibitem[Lee et al.(2005)]{Lee05} Lee, Y.-W., et al.\ 2005, \apjl, 621, L57 
\bibitem[Lee et al.(2007)]{2007ApJ...661L..49L} Lee, Y.-W., Gim, H.~B., 
\& Casetti-Dinescu, D.~I.\ 2007, \apjl, 661, L49
\bibitem[Norris(2004)]{2004ApJ...612L..25N} Norris, J.~E.\ 2004, \apjl, 
612, L25
\bibitem[O'Connell(1999)]{1999ARA&A..37..603O} O'Connell, R.~W.\ 1999, \araa, 37, 603

\bibitem[Park \& Lee(1997)]                    {Park97}Park, J.-H., \& Lee, Y.-W. 1997, \apj, 476, 28
\bibitem[Piotto et al.(2005)]{2005ApJ...621..777P} Piotto, G., et al.\ 
2005, \apj, 621, 777 
\bibitem[Piotto et al.(2007)]{2007ApJ...661L..53P} Piotto, G., et al.\ 2007, \apjl, 661, L53 
\bibitem[Ree et al.(2007)]{Ree07} Ree, C.~H., et al.\ 2007, 
\apjs, 173, 607 
\bibitem[Reimers(1977)]{1977A&A....57..395R} Reimers, D.\ 1977, \aap, 57, 395
\bibitem[Rose(1985)]{1985AJ.....90.1927R} Rose, J.~A.\ 1985, \aj, 90, 1927
\bibitem[Schiavon et al.(2006)]{2006ApJ...651L..93S} Schiavon, R.~P., et 
al.\ 2006, \apjl, 651, L93
\bibitem[Schiavon(2007)]                {Schi07}Schiavon, R.~P. 2007, \apjs, 171, 146
\bibitem[Siegel et al.(2007)]{2007ApJ...667L..57S} Siegel, M.~H., et al.\ 
2007, \apjl, 667, L57
\bibitem[Sohn et al.(2006)]{2006AJ....131..866S} Sohn, S.~T., O'Connell, 
R.~W., Kundu, A., Landsman, W.~B., Burstein, D., Bohlin, R.~C., Frogel, 
J.~A., \& Rose, J.~A.\ 2006, \aj, 131, 866
\bibitem[Thomas et al.(2005)]{2005ApJ...621..673T} Thomas, D., Maraston, 
C., Bender, R., \& Mendes de Oliveira, C.\ 2005, \apj, 621, 673
\bibitem[Trager et al.(2000)]                  {Trag00}Trager, S. C., Faber, S. M., Worthey G., Gonzalez J. J. 2000, \aj, 119, 1645
\bibitem[Trager et al.(2008)]{2008MNRAS.386..715T} Trager, S.~C., Faber, 
S.~M., \& Dressler, A.\ 2008, \mnras, 386, 715
\bibitem[Ventura 
\& D'Antona(2008)]{2008MNRAS.385.2034V} Ventura, P., \& D'Antona, F.\ 2008, \mnras, 385, 2034
\bibitem[Worthey et al.(1992)]{1992ApJ...398...69W} Worthey, G., Faber, 
S.~M., \& Gonzalez, J.~J.\ 1992, \apj, 398, 69
\bibitem[Yi et al.(1998)]{1998ApJ...492..480Y} Yi, S., Demarque, P., 
\& Oemler, A., Jr.\ 1998, \apj, 492, 480
\bibitem[Yi et al.(1999)]{1999ApJ...513..128Y} Yi, S., Lee, Y.-W., Woo, 
J.-H., Park, J.-H., Demarque, P., \& Oemler, A., Jr.\ 1999, \apj, 513, 128
\bibitem[Yoon et al.(2008)]{Yoon08} Yoon, S.-J., Joo, S.-J., Ree, C.~H., Han, S.-I., Kim, D.-G., \& Lee, Y.-W.\ 2008, \apj, 677, 1080

\end{thebibliography}
\end{document}